\documentclass[9.5pt,alternate,final,finalsubmission,twocolumn]{sig-alternate-10pt}

\usepackage{epsfig}
\usepackage{latexsym}
\usepackage{epsf}
\usepackage{graphicx}

\newcommand{\doctitle}{A Study of Geolocation Databases}

\newcommand{\rem}[1]{}

\newcommand{\ignore}[1]{}
\newcommand{\etal}{{\em et al.}}
\newcommand{\parent}{\emph{parent pair }}
\newcommand{\child}{\emph{child pair }}

\def\FullBox{\hbox{\vrule width 8pt height 8pt depth 0pt}}
\newcommand{\qqqed}{\;\;\;\FullBox}

\newcommand{\ls}[1]
   {\dimen0=\fontdimen6\the\font
    \lineskip=#1\dimen0
    \advance\lineskip.5\fontdimen5\the\font
    \advance\lineskip-\dimen0
    \lineskiplimit=.9\lineskip
    \baselineskip=\lineskip
    \advance\baselineskip\dimen0
    \normallineskip\lineskip
    \normallineskiplimit\lineskiplimit
    \normalbaselineskip\baselineskip
    \ignorespaces
   }
\newcounter{LineNum}
\newcounter{AlgNum}

\newsavebox{\savepar}

\begin{document}

\title{\doctitle}

\numberofauthors{2}
\author{
\alignauthor
Yuval Shavitt\\
\affaddr{School of Electrical Engineering}\\
\affaddr{Tel-Aviv University, Israel}\\
\email{shavitt@eng.tau.ac.il} \alignauthor
Noa Zilberman\\
\affaddr{School of Electrical Engineering}\\
\affaddr{Tel-Aviv University, Israel}\\
\email{noa@eng.tau.ac.il} }
 \maketitle

\begin{abstract}

The geographical location of Internet IP addresses has an importance
both for academic research and commercial applications. Thus, both
commercial and academic databases and tools are available for
mapping IP addresses to geographic locations.

Evaluating the accuracy of these mapping services is complex since
obtaining diverse large scale ground truth is very hard. In this
work we evaluate mapping services using an algorithm that groups IP
addresses to PoPs, based on structure and delay. This way we are
able to group close to 100,000 IP addresses world wide into groups
that are known to share a geo-location with high confidence. We
provide insight into the strength and weaknesses of IP geolocation
databases, and discuss their accuracy and encountered anomalies.

\end{abstract}
\footnotetext[2]{This work was partially funded by the  OneLab II
and the MOMENT consortia, which are partly financed by the European
Commission; and by the Israeli Science Foundation, center of
knowledge on communication networks (grant \#1685/07)}

\section{Introduction}\label{intro.sec}

Geolocation services have become in the recent years a necessity in
many fields and for many applications. While the end user is usually
not aware of it, many websites visited by him every day use
geolocation information. Some of the common uses of geolocation
information is for targeted localized advertising, localized content
(such as local news and weather), and compliance with local law.

Perhaps the most highlighted purpose of geolocation information is
for fraud prevention and various means of security. Banking,
trading, and almost any other type of business that handles online
money transactions are exposed to phishing attempts as well as other
schemes. Criminals try to break into user accounts to transfer
money, manipulate stocks, make purchases and more. The geolocation
information provides means to reduce the risk, for example by
blocking users from certain high-risk countries, cross-referencing
user expected and actual location and more. Organizations that
handle national security find geolocation information useful as
well, like the DHS cyber security center \cite{DHSCS09}. Even simple
emergency services, such as dispatching emergency responders to the
location of emergency use it.

Geolocation information is also important in many research fields.
It improves internet mapping and
characterization, as it ties the internet graph to actual node
positions, and allows exploring new aspects of the network that are
otherwise uncovered, such as the effect of ISP location on its
services and types of relationships with other service providers.

Many previous papers have discussed the usage of geolocation
information in day-to-day applications. They vary in fields from law
\cite{JE06,SV07,SV08} through information security and fraud
prevention \cite{Ma09,GS07} to various economic aspects
\cite{KF10,King09}. However, not many works have focused on the
accuracy of geolocation databases. In 2008, Siwpersad
\etal~\cite{SGU08} examined the accuracy of Maxmind~\cite{maxmind}
and IP2Location~\cite{ip2location}. They assessed their resolution
and confidence area and concluded that their resolution is too
coarse and that active measurements provide a more accurate
alternative. Gueye \etal~\cite{GUF07} investigated the imprecision
of relying on the location of blocks of IP addresses to locate
Internet hosts and showed that the geographic area spanned by blocks
can be far larger than the typical distance between any two IPs
within a block. Thus it indicated that geolocation information
coming from exhaustive tabulation may contain an implicit
imprecision.

The IETF has also commenced in defining standards for geolocation
and emergency calling. The IETF GEOPRIG working group
\cite{IETF-Geo} discusses internet geolocation standards and privacy
protection for geolocation. Some examples are DHCP location, as in
RFC3825 and RFC4776, and defining protocols for discovering the
local location information server \cite{IETF-LIS}.

Muir and Oorschot \cite{MO09} conducted a survey of geolocation
techniques used by geolocation databases and examined means for
evasion/circumention from a security standpoint.

Improving location accuracy by measurements has been addressed by
several works in the recent years. IP2Geo \cite{PS01} is one of the
first to suggest a measurement-based approach to approximate the
geographical distance of network hosts. A more mature approach is
constraint based geolocation \cite{GZCF06}, which uses several delay
constraints to infer the location of a network host by a
triangulation-like method. Later works, such as Octant
\cite{WongSS07}\, use a geometric approach to localize a node within
22 mile radii. Katz-Bassett \etal~\cite{Katz06} suggested topology
based geolocation using link delay to improve the location of nodes.
Yoshida \etal~\cite{Yoshida09} used end-to-end communication delay
measurements to infer PoP level topology between thirteen cities in
Japan. Laki \etal~\cite{LMHCV09} increased geolocation accuracy by
decomposing the overall path-wise packet delay to link-wise
components and were thus able to approximate the overall propagation
delay along the measurement path. Eriksson \etal~\cite{EBSN10} apply
a learning based approach to improve geolocation. They reduce IP
geolocation to a machine learning classification problem and use
Naive Bayes framework to increase geolocation accuracy.

In this paper we study the accuracy of geolocation databases. The
main problem in such a study is the lack of ground truth
information, namely a large and diverse set of IP addresses with
known geographic location to compare the geolocation databases
against.  We avoid this need using a different approach, we use an
algorithm, whose main features are summarized in
Section~\ref{model.sec}, for mapping IP addresses to PoPs (Points of
Presence).  The algorithm, which is based both on delay measurements
and graph structure, has a very small probability to map two IP
addresses, which are not co-located, to the same PoP.  Thus, while
we do not know the location of the PoP we know that all the IP
addresses within a PoP should reside in the same location. This
serves as a mean to check a geolocation database coherency: if two
IP addresses in the same PoP are mapped to different locations the
database has a problem, and we can use the distances among the
various locations of IP addresses in the same PoP as a measure of
database accuracy.  The results are presented in
Sec.~\ref{sec:basic-tests}.

We take a step further and compare multiple databases results for
the same PoP (Sec.~\ref{subsec:compare}). In case a majority of the
results in each database are identical we can expect that for each
database a majority vote will give us the correct location of the
PoP, and the spread of the locations will give us the confidence
measure of the result.  This will help us to identify cases where a
database reports for a large portion of the IP addresses of some ISP
the same default location (usually the ISP headquarters).

\section{GeoLocation Services}\label{geolocation.sec}
Geolocation services range from free services, through services that
cost a few hundreds of dollars and up to services that cost tens of
thousands of dollar a year. This section surveys most of these
services,  focusing on the main players.

Free geolocation services differ from one another in nature. Three
representative of such sources are discussed here: DNS resolution,
Google Gears and HostIP\-.Info. DNS resolution was probably the
first source for geolocation information, being free and available
to all users. In 2002 Spring \etal~\cite{Spring02} used DNS names to
improve location information as part of the Rocketfuel project. The
UnDNS tool they provided is still used to uncover location from DNS
name. However, DNS suffers from several problems: many interfaces do
not have a DNS name assigned to them, and incorrect locations are
inferred when interfaces are misnamed \cite{ZRPR06}. In addition,
rules for inferring the locations of all DNS names do not exist, and
require some manual adjustments. As part of Google Labs Gears API,
Google provides a set of geolocation API~\cite{google} that allows
to query a user's current position. The position is obtained from
onboard sources, such as GPS, a network location provider, or from
the user's manual input. When needed, the location API also has the
ability to send various signals that the devices has access to
(nearby cell sites, WIFI nodes, etc.) to a third-party location
service provider, who resolves the signals into a location estimate
\cite{google2}. Thus, the service granularity is based on a single
IP address granularity and not on address blocks.
HostIP.Info~\cite{hostip} is an open source project, with many of
its API contributed by its community. The data is collected from
users participating in direct feedback through the API, as well as
ISP's feedback. In addition, website visitors are updating their
location, which in turn is updated as a database entry. The city
data comes from various sources, such as data donation and US census
data (for the USA). The data is provide as $/24$ CIDR blocks.

Another type of geolocation services emerges from universities and
research institutes. These services tend to use measurements,
entirely or on top of other methodologies, in order to improve
geolocation data quality. While many of the measurement based
geolocation services that we discussed in Section \ref{intro.sec} do
not provide the ability to query specific IP addresses
\cite{Katz06,WongSS07,Yoshida09}, one online geolocation service
that does allow it is Spotter~\cite{spotter}, which is based on a
work by Laki \etal~\cite{LMHCV09}.  Spotter uses a detailed
path-latency model to determine the overall propagation delays along
the network paths more accurately, which in turn translates to more
accurate geographic distance estimation. The evaluation process also
takes into account the discovered topology between the measurement
points, and end-to-end latency measurements as well. One-way delay
measurements further increase the accuracy of router geolocation
techniques.

Mid-range cost geolocation services include databases such as
Maxmind GeoIP, IPligence, and IP2Location. All these databases cost
a few hundreds of US Dollars and provide the user a full database,
typically as a flat file or MySQL dump. Some of the companies, such
as MaxMind, also provide a geolocation web service.

MaxMind \cite{maxmind} is one of the pioneers in geolocation,
founded in 2002, and it provides a range of databases: from country
level to city level, longitude and latitude. Information on ISP and
netspeed can be retrieved as well. In addition to all the above,
MaxMind suggests to enterprises a database with an accuracy radius
for its geolocation information. In this work, the MaxMind GeoIP
City database is being used for geolocation information.
IPInfoDB~\cite{ipinfo} is a free geolocation service that uses
MaxMind GeoIP lite database and adds on top of it reserved addresses
and  optional timezone.

IPligence \cite{ipligence} is a geolocation service provider,
existing since 2006. Its high end product, IPligence Max, provides
geographic information such as country, region and city, longitude
and latitude, in addition to general information such as owner and
timezone. Hexasoft development maintains
IP2Location~\cite{ip2location}, a gelocation database with a wider
range of geolocation information: from IP to country conversion, to
retrieving information such as bandwidth and weather. For this
study, we used their DB5 database, which maps IP addresses to
country, region, city, latitude, and longitude. In all the above
products, the IP addresses' location is given in ranges, which vary
in size and reach the granularity of a handful of addresses per
range.

High end geolocation services are often priced by the number of
queries and their cost may reach tens of thousands of dollars a year
for large websites. Amongst these services, and based on their
pricing level, are Quova, Akamai Edge Platform, Digital Element's
Netacuity Edge and Goebytes. Each of these companies praise
themselves with large tier-1 customers from different fields, who
use their services for target advertising, fraud prevention, and
more.

Quova~\cite{quova}, founded in 1999, provides three levels of data
information, bronze, silver, and gold. The advanced services contain
attributes such as location confidence level, Designated Market Area
(DMA), and status designations for anonymized Internet connections.
Quova's database is based on random forest classifier rules,
synthesis rules, approved location labels, hand-labeled hostnames,
and research notes, with 6 patents issued and 9 patents pending.

Akamai~\cite{akamai} was founded in 1998 and lunched its commercial
service in 1999, provides through its Edge Platform product IP
location information. Its IP location services are a part of a much
larger package of tools and applications used for traffic
management, dynamic sites accelerations, performance enhancement and
more.

Digital Element~\cite{netacuity}, founded in 2005, provides under
the products NetAcuity and NetAcuity Edge two levels of geolocation
information, with over thirty nine data points, including
demographics, postal code, and business type. The IP geolocation
data source is anonymous data gathered from interactions with users.
One source for this user information is partner companies that use
the product. The information is validated using a proprietary
clustering analysis algorithm. The data collection and analysis are
protected by more than 20 issued and pending patents.

Geobytes~\cite{geobytes}
launched in 2002 its GeoSelect product, which provides geolocation
information. The data provided by Geobytes matches mid-range
companies in its wealth, but it is part of a broader package of
services, including reports, users redirection, etc. While in the
past Geobytes used ICMP packets to create an infrastructure map,
current methods include also gathering information from websites
that require users to enter their location information and then
processing this data onto Geobytes' infrastructure map of the
Internet \cite{geobytes_patent}. No DNS information is used by
Geobytes for their location resolution.

In this work, databases from all three groups are being used. From
the no-charge databases: HostIP.info, Spotter and DNS (partial).
Mid-range databases used are MaxMind GeoIP City, IPligence Max, and
IP2\-Location DB5. GeoBytes and NetAcuity are the last two databases
used in this work. Unfortunately, we failed to reach a collaboration
with Quova and Akamai for this project.

\subsection{Databases Accuracy}
The geolocation service provider is, in many cases, the sole source
for database accuracy information. Some vendors do not publish
accuracy figures at all, such as IPligence, while others provide
accuracy figures without explaining how they were obtained. A few
geolocation services, such as Akamai and Quova provide accuracy
figures obtained by external auditors. Table \ref{tab:accuracy}
provides a summary of accuracy figures, as given by the geolocation
service providers on their websites
\cite{quova,akamai,netacuity,geobytes,maxmind,ip2location}. The
table includes information on country level accuracy, city level
accuracy world wide and city level accuracy in the USA.
\begin{table}
\begin{minipage}[b]{\linewidth}
\begin{center}
\small\addtolength{\tabcolsep}{-3pt}

\begin{tabular}{|l|c|c|c|}
 \hline
    \bf{Database} & {\bf{Country Level }} & {\bf{City Level }} & {\bf{USA City Level}}\\
\hline
     IP2Location & 99\%&  80\% & \\
\hline
     MaxMind & 99.8\% & Varies & 83\% \\
\hline
     GeoBytes & 97\%&  85\% &  \\
\hline
     NetAcuity & 99.9\% & 95\% &  \\
\hline
     Akamai & & 97.22\% & 100\% \\
\hline
     Quova & 99.9\% &  & 97.2\% \footnotemark[2] \\
\hline

\end{tabular}
\caption{Geolocation Database Accuracy } \label{tab:accuracy}
\end{center}
\end{minipage}
\end{table}

\footnotetext[2]{State level accuracy}

All the databases claim to have $97\%$ accuracy or more at the
country level and $80\%$ or more at the city level. MaxMind provides
detailed expected accuracy on city level based on country
\cite{maxmind_city}. The accuracy ranges from $40\%-44\%$ in
countries like Nigeria and Tunisia to $94\%-95\%$ in countries like
Georgia, Qatar and Singapore. An accurate resolution here is
considered one within 25 miles from its true location.  Netacuity
accuracy figures are based on a test by Keynote Systems, which
resulted in an exact match for every IP address at the country
level, and state level for those IP addresses located in the United
States. At the city level NetAcuity delivered 97\% accuracy. Quova's
accuracy figures are based on an audit by Pricewaterhouse Coopers
\cite{quova_audit}, which used 3 reference third party databases.
Here 99.9\% accuracy was achieved at the country level and 97.2\% to
98.2\% were achieved at the state level.

The accuracy of the figures in Table \ref{tab:accuracy} cannot be
easily evaluated. For example, neither the means by which Keynote
Systems tested Netacuity nor the reference databases used to test
Quova are revealed. Akamai claims for 97.2\% accuracy at the city
level worldwide and 100\% accuracy at the city level in the USA. The
source for this is a report by Gomez \cite{akamai_audit}, which
defined a node location to be unique on /23 CIDR subnets. In
addition, a Census Metropolitan Area (CMA) is the basis of the
naming convention used by Gomez to identify the physical location of
its measurement nodes. The accuracy of this method is thus
debateable, as described in Sec.~\ref{intro.sec}.

Assesing the accuracy of geolocation databases is therefore a hard
question, since a large scale ground truth does not exist (or is
hard to obtain). In this work a structural approach is taken to
evaluate the accuracy of databases, gaining greater knowledge on
each IP address by locating it as part of a PoP-level map.

\section{The Evaluation Model}\label{evaluation.sec}

\subsection{Building PoP Maps}\label{model.sec}
We define a PoP as a group of routers which belong to a single AS
and are physically located at the same building or campus. In most
cases \cite{JUNIPOP,CiscoPoP} the PoP consists of two or more
backbone/core routers and a number of client/access routers. The
client/access routers are connected redundantly to more than one
core router, while core routers are connected to the core network of
the ISP. The algorithm we use for PoP extraction was first suggested
by Feldman and Shavitt in \cite{FeldmanS08} and later improved in
\cite{SZ-Netsci10}. The algorithm looks for bi-partite subgraphs
with certain weight constraints in the IP interface graph of an AS;
no aliasing to routers is needed. The bi-partites serve as cores of
the PoPs and are extended with other close by interfaces.

The initial partitioning removes all edges with delay higher than
$PD_{max\_th}$, PoP maximal diameter threshold, and edges with
number of measurements below $PM_{min\_th}$, the PoP measurements
threshold. $PM_{min\_th}$ is introduced in order to consider only
links with a high reliable delay estimation to avoid false
indication of PoPs. The result non-connected graph $G'$ contains
induced sub graphs, each is a candidate to become one or more PoPs.
There are two reasons for a connected group to include more than a
single PoP. The first and most obvious reason is geographically
adjacent PoPs, e.g., New York, NY and Newark, NJ. The other is
caused by wrong delay estimation of a small amount of links. For
instance a single incorrectly estimated link between Los Angeles,CA
and Dallas,TX might unify the groups obtained by such a naive
method.

Next, the algorithm checks if each connected group can be partitioned to more than one
PoP, using parent-child classification according to the measurement
direction in the bipartite graph. Further localization is achieved
by dividing the parents and children groups into physical
collocations using the high connectivity of the bipartite graph. If
\parent and \child groups are connected, then the weighted distance
between the groups  is calculated (If they are connected, by
definition more than one edge connects the two groups); if it is
smaller than a certain threshold the pair of groups is declared as
part of the same PoP. Last, a unification of loosly connected parts
of the PoP is conducted. For this end, the algorithm looks for
connected components (PoP candidates) that are connected by links
whose median distance is very short (below $PD_{max\_th}$).

In the original algorithm \cite{FeldmanS08}, an additional step was
implemented, called Singleton Treatment, in which nodes with only
one or two links are assigned to PoPs based on their median
distance. This step may add to the PoP IP addresses that are not
necessarily part of it. Thus, in this work, two PoP level maps were
generated: one map without any singletons, which is considered to be
accurate looking at the PoP IP addresses only, and a second map
that includes singletons. The aim of the second map is to improve location
estimation where PoP location is undetermined based on the first map
only. As the singletons are necessarily in the vicinity of the PoP,
using them does not harm the locations estimation.

\begin{figure}
\begin{minipage}[b]{\linewidth}
\centering
\includegraphics[width=8cm]{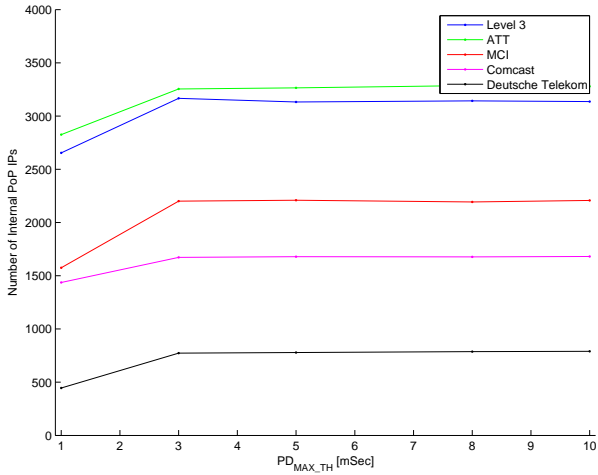}
\caption{Number of IPs in PoPs vs. Maximal Delay}
\label{fig:delay_sensitivity_ips}
\end{minipage}
\end{figure}

In a previous work~\cite{SZ-Netsci10},  the stability and
correctness of the PoP extraction algorithm were discussed, as well
as the effect of threshold settings. For this paper's purposes, the
thresholds sensitivity should be mentioned, as they may affect the
geolocation accuracy. Figure \ref{fig:delay_sensitivity_ips}
explores the PoP extraction algorithm's sensitivity to
$PD_{max\_th}$. In the figure five ISPs are explored: Level 3,
AT\&T, Comcast, MCI, and Deutsche Telekom. The figure presents the
number of IPs included in PoPs when changing $PD_{max\_th}$. Neither
the number of discovered PoPs nor the number of IPs within the PoPs
are sensitive to the delay threshold, as long as the threshold is
$3mSec$ or above. $PD_{max\_th}$ was selected to be $5mSec$, as it
presents a good tradeoff between delay measurement's error and
location accuracy. The number of IPs included in PoPs decreases as
the minimal number of required measurements, $PM_{min\_th}$,
increases, as can be expected (see \cite{SZ-Netsci10}). In our
extracted PoP maps, $PM_{min\_th}$ was selected to be $5$.

\subsection{Data Evaluation Method}
The geolocation databases evaluation is conducted using the
classification of IP addresses into PoPs as described above.  Since
the classification is based on both structure and delay
measurements, the chances that two IP addresses, which our algorithm
map to the same PoP, are not located in the same geographical
location are slim.  We do recognize that when two PoPs are very
close (within a few tens of kilometers) our algorithm may unify them
to one.  However, in this case the median of their location is half
their distance, namely not far.

To identify the geographical location of a PoP, we use
the geographic location of
each of the IPs included in it. As all the PoP IP addresses should
be located within the same campus, or within its vicinity if
singletons are considered, the location confidence of a PoP is significantly
higher than the confidence that can be gained from locating each of
its IP addresses separately. The algorithm, introduced in
\cite{SZ-Netsci10}, operates as follows:

{\bf Initial Location } Each of the evaluated geolocation databases
is queried for the location (longitude, latitude) of each IP
included in the PoP. Next, the center weight of the PoP location is
found by calculating the median of all PoP's IP locations. Unlike
average calculation, where a single wrong IP can significantly
deflect a location, median provides a better suited starting point.
Median is certainly not a guarantee for good results. If there is
complete disagreement between geolocation databases as for the
location of a PoP, e.g., if one of them places all the PoP IPs in
London, and the other in New-York, the median may be far away from
any of the suggested locations. However, since geolocation databases
are typically reliable in country-level assignment, such an example
is highly unlikely. We consider this assumption later in section
\ref{results.sec}.

{\bf Location Error Range } Every PoP location is assigned a range
of convergence, representing the expected location error range based
on the information received from the geolocation databases. As the
PoP location is given as [latitude, longitude], in units of degrees,
so does the range of convergence. This stage is done iteratively,
looking for a majority vote for the PoP location. For every IP
address in a PoP and for every geolocation database we collect the
geographic coordinates, thus if there are $N$ IP addresses and $M$
databases, and for each of the IP addresses, all the databases
suggest a location, then $N\times M$ IP address elements are being
considered for the vote. The algorithm starts at the median
location, and checks if there is a majority vote for the PoP
location within a radius 0.01 degrees (one latitude/longitude degree
is roughly equivalent to 111km). If the circle includes less than
50\% of the located IP elements, we continue and increase the radius
of the circle, by 0.01 degrees each step, until the PoP location has
a majority vote. Alternatively, the algorithm stops when the circle
radius reaches a predefined threshold, typically 1 or 5 degrees,
which we define as the maximal range of error. If one of the
geolocation databases lacks information on an IP address, this IP
element is not counted in the majority vote. With a majority vote we
ensure most of the geolocation databases agree on the PoP location.

{\bf Location Refinement } After a range of convergence is found,
the PoP location accuracy is further improved. A new median location
of the PoP is calculated, based only on IP elements that are located
within the range of convergence. This ensures that deviations in the
PoP location caused by a small number of IP elements outside the
range of convergence are discarded, and the PoP is
centered based only on credible IP addresses.

The result of the PoP geolocation algorithm includes per PoP the
following new parameters: longitude, latitude, range of convergence,
the percentage of IP addresses within convergence range out of all IP
addresses, and the percentage of IP addresses within the covergence
range considering only IP elements with location information.

To validate the PoP geolocation generated maps correctness, results
were compared against PoP maps published by the ISPs, such as Sprint
\cite{sprint_map}, Qwest \cite{qwest_map}, Global Crossing
\cite{gblx_map}, British Telecom \cite{british_map}, AT\&T
\cite{att_map} and others. In addition, we reported
\cite{SZ-Netsci10} a limited small scale testing of the geolocation accuracy based on
50 known university locations. The test was based
only on three databases: Maxmind, IPligence and HostIP.info.
For 49 out of 50
universities, the location was accurate within a 10 kilometer
radius. The last PoP, belonging to The University of Pisa, was
located by the algorithm in Rome, due to an inaccuracy in the
MaxMind and Ipligence databases. Only Hostip.info provided the right
coordinates for this PoP. Each PoP location was also validated
against its DNS name, whenever a DNS name was assigned to the
interface.

\subsection{Dataset}
\begin{figure}
\begin{minipage}[b]{\linewidth}
\centering
\includegraphics[width=8cm, height=4cm]{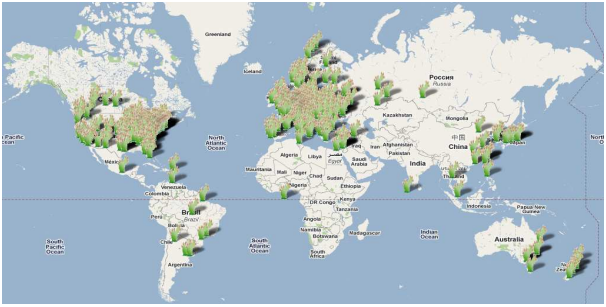}
\caption{Map Of DIMES Agents, March-2010} \label{fig:agents_map}
\end{minipage}
\end{figure}

The collected dataset for PoP level maps is taken from
DIMES~\cite{DimesSigcomm05}. We use all traceroute measurements
taken during March 2010, totaling $126.7$ million, namely an average
of 4.2M million measurements a day. The measurements were collected
from over $1750$ vantage points, which are located in $74$ countries
around the world, as shown in Figure \ref{fig:agents_map}. About
$16\%$ of the vantage points are mobile.

The $126.7$ million measurements produced $7.85$ million distinct IP level
edges (no IP level aliasing was performed).
Out of these, $1.3$ million edges were measured five
times or more, thus above $PM_{min\_th}$, and $642K$ edges had less
than $PD_{max\_th}$ median delay, and were therefore considered by
the PoP extraction algorithm. As described above, two PoP level maps were generated by
the PoP extraction algorithm, with and without singletons addition.
A total of $3800$ PoPs where
discovered, containing $52K$ IP addresses from the first run, and
$104K$ IP addresses from the second run, meaning with singletons.
Although the number of discovered PoPs is not large, as the
algorithm currently tends to discover mainly large PoPs while missing many
access PoPs, the large number of IP addresses and the spread around
the world (see below) allow a large scale and meaningful geolocation databases evaluation.

\begin{figure}
\begin{minipage}[b]{\linewidth}
\centering
\includegraphics[width=8cm, height=4cm]{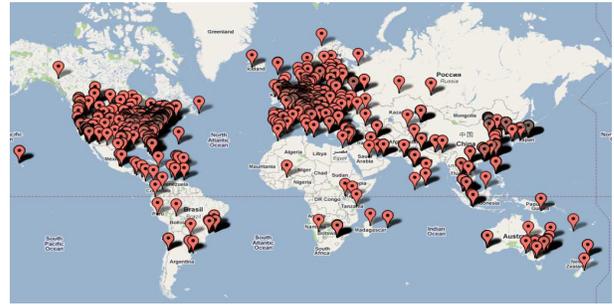}
\caption{Map Of Discovered PoPs, March-2010} \label{fig:pops_map}
\end{minipage}
\end{figure}

Figure \ref{fig:pops_map} shows the geographical location (as
calculated by our algorithm) of the PoPs discovered by the PoP
algorithm. The PoPs are spread all over the world, in all five
continents, with high density of PoPs in Europe and North America.
As can be seen, PoPs are located even in places such as Madagascar
and Papua New Guinea, which comes to show the vast range of location
information required from the geolocation databases in this
evaluation.

The following databases were studied in this work: MaxMind GeoIP,
IPligence Max, IP2Location DB5, Ho\-stIP.info, GeoBytes, NetAcuity,
DNS and Spotter. For most of the databases, the data which was used,
was updated on the first week of April 2010. NetAcuity database was
obtained on the third week of April and Spotter located the IP
addresses during April and the beginning of May 2010.

\section{Results}\label{results.sec}
\subsection{Basic Tests}  \label{sec:basic-tests}

\subsubsection{Null Replies}

The first question asked for each database is "How many NULL replies
are returned for IP address queries?". There are four flavors for
this question. First it is asked only on IPs which are in the core
of the PoPs and then it is asked for all IP addresses, including
singletons addresses. As some databases may have better information
on end users or access interfaces than on core routers and main
PoPs, this can be meaningful. The next observation regards NULL
replies that apply to all the IP addresses within a certain PoP:
does the database fail to cover a range of addresses or a physical
location range, or are the NULL replies a matter of a single IP
address lack of information? This too is considered both with and
without singletons. Table \ref{tab:nulldata} shows for each of the
databases the percentage of IP addresses which returned a NULL reply
for each of these questions.

\begin{table}
\begin{minipage}[b]{\linewidth}
\begin{center}
\small\addtolength{\tabcolsep}{-3pt}

\begin{tabular}{|l|c|c|c|c|}
 \hline   \bf{} & \multicolumn{2}{c|}{\bf{Core PoP IP }} & \multicolumn{2}{c|}{\bf{With Singletons
   }}\\
 \hline
    \bf{Database} & {\bf{Null IP }} & {\bf{Null PoP }} & {\bf{Null IP }} & {\bf{Null PoP}}\\
 \hline
     IPligence & 3.9\%&  1.5\% & 2.9\% & 1.4\% \\
\hline
     IP2Location & 0\%&  0\% & 0\% & 0\% \\
\hline
     MaxMind & 36\%&  10.6\% & 30.1\% & 6\% \\
\hline
     HostIP.Info & 64\%&  38.6\% & 64\% & 29\% \\
\hline
     GeoBytes & 20.7\%&  4.3\% & 17.8\% & 2.7\% \\
\hline
     NetAcuity & 0\%&  0\% & 0\% & 0\% \\
\hline
     Spotter & 37\%&  18.1\% &   &   \\
\hline
     DNS & 14.3\%& 12.2\% & 28.4\% &  2\%\\
\hline

\end{tabular}
\caption{Null IP Address Information } \label{tab:nulldata}
\end{center}
\end{minipage}
\end{table}

NetAcuity and IP2Location where the only databases to return a reply
for all the queried IP addresses. For IP2Location database there are
a few hundreds of NULL entries in the entire database (for IP
addresses not in this study). This alone does not come to indicate
that the returned addresses are correct, only that an entry exists.
The location correctness is discussed later on in this section. On
the other end of the scale, HostIP.info failed to locate most of the
IP addresses, however on the PoP level this percentage drops by
half. It can be assumed that HostIP.info nature of the failure is
lack of information on specific IP addresses and not IP ranges.
Further more, in most cases HostIP.info does return a reply with
country information, but without longitude and latitude. Spotter did
not locate about a third of the IP addresses. The reason for such a
failure can be either that the IP did not respond to ping or the IP
responded to ping, but the roundtrip-times were too high to provide
approximations for the algorithm. Only core PoP IP addresses,
without singletons, where tested here. For MaxMind, the percentage
of Null replies refers to events where no specific location
information was available. In most of these cases, MaxMind does
return longitude and latitude information, which are the center of
the country where the IP is located. A list of these coordinates is
available to the users, and though we choose in this work to refer
to this information as a NULL reply, a general notion of location is
provided by the database. DNS NULL replies are less than 15\% for
core PoP IP adresses, and almost 29\% when taking into account
singletons. As there is a probability that singletons represent end
users and not router interfaces, this is expected. The effect of
grouping to PoPs when looking at DNS is significant: when taking
into account singletons, only 2\% of the PoPs have no location by
DNS.

\subsubsection{Agreement within database}
By nature, IP addresses belonging to the same PoP reside in the same
area. One can leverage this information to evaluate the accuracy of
a geolocation database: if IP addresses that belong to the same PoP
are assigned different geographical location, then the accuracy of
this information should be questioned. This statement is based on
the assumption that the PoP algorithm is correct and does not assign
IP addresses from different locations to the same PoP. We
already discussed why it is true based on design and previous
limited evaluation.  Our experiments here further support the assumption:
in all the PoPs evaluated, with no
exception, there are always databases that support the PoP vicinity
requirement.

\begin{figure}
\begin{minipage}[b]{\linewidth}
\centering
\includegraphics[width=8cm, height=5.5cm]{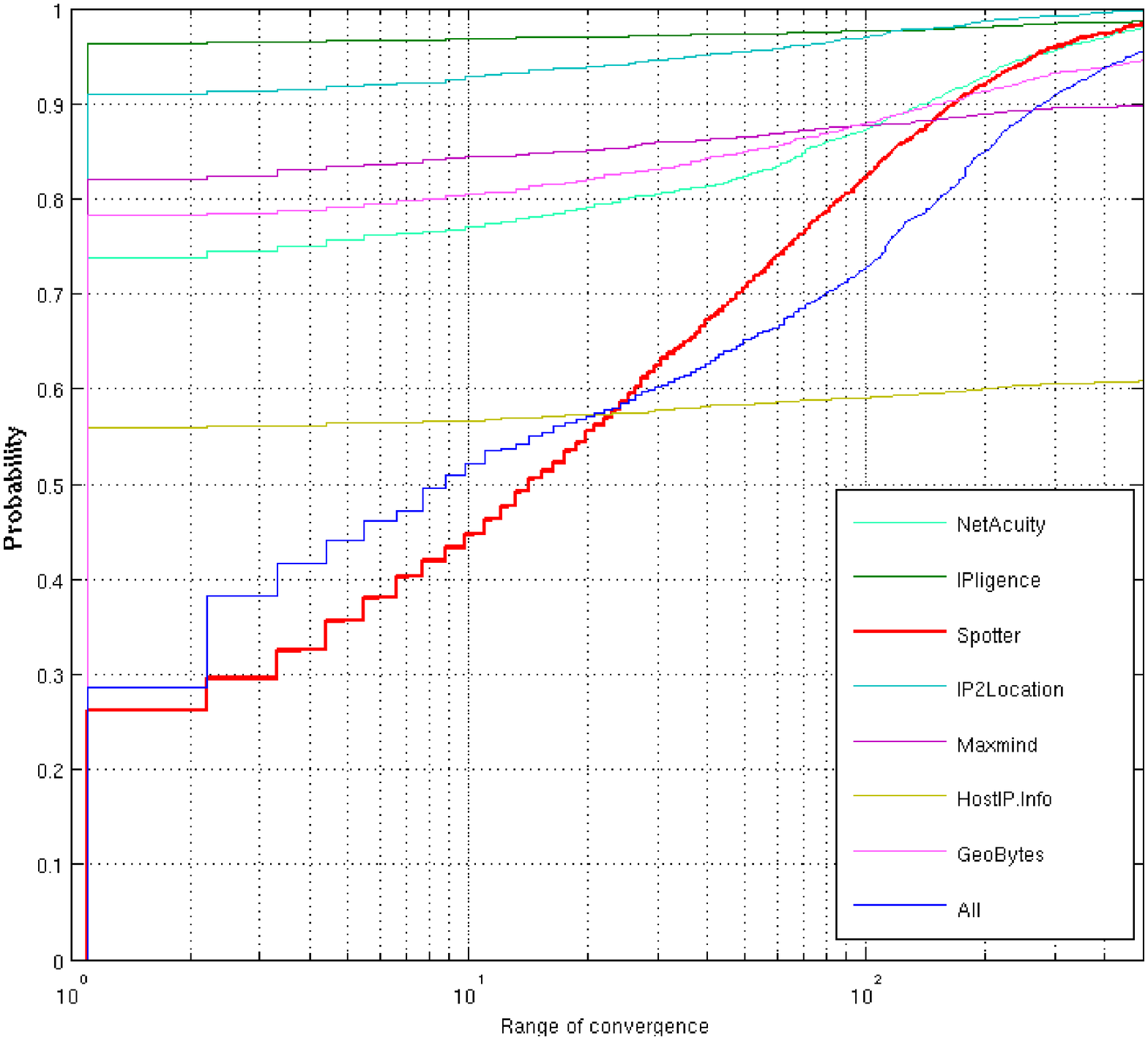}
\caption{Range of Convergence Within Databases}
\label{fig:cdf_range}
\end{minipage}
\end{figure}

Figure \ref{fig:cdf_range} presents a CDF of the convergence range
within databases without singletons. The X-axis is the range of
convergence in kilometers, logarithmic scale, with 500km being the
limit where the algorithm was stopped. The algorithm progressed its
testing in steps of 1km.

IPligence and IP2Location clearly have a range of convergence far
better than other databases: over $90\%$ of the PoPs located using
these databases have the minimal range of convergence - one
kilometer, which is in practice the exact same location. MaxMind,
GeoBytes and NetAcuity have $74\%$ to $82\%$ of their PoPs converge
within one kilometer. For HostIP.Info, a bit less than $57\%$ of the
PoPs converge within the minimal range, and almost all the rest fail
to converge. This is caused mostly due to lack of information on IP
addresses, as many PoPs do not have even a single IP with location
information inside a PoP. The case of Spotter here is different. As
this information is acquired by measurements, having almost a third
of the PoPs converge within one kilometer is an indication of good
performance. In addition, over $82\%$ of the PoPs converge within
$100km$, and close to 98\% within $500km$, which is similar or
better than most of the other databases. The slow accumulation is
expected due to measurements errors. Maybe the most important graph
here is the $All$ graph, showing the range of convergence when
combining the information from all databases. Though all databases,
have most of their PoPs located within the minimal range, less than
$30\%$ of the $All$ PoPs converge within this range, meaning that
between the databases there is disagreement, though as the range
grows so does the percentage of converged PoPs. This does not
necessarily mean that all the databases have agreed on the same
location, as databases which reply with a location for every IP have
more influence that databases with some NULL replies. We further
explore this question in section \ref{subsec:compare}. An important
observation is that even if a certain database indicates that the
range of convergence of a PoP is minimal, i.e., 1km, it does not
necessarily imply accuracy, or in our case that all other databases
will agree with this location.

\begin{figure}
\begin{minipage}[b]{\linewidth}
\centering
\includegraphics[width=8cm, height=5.5cm]{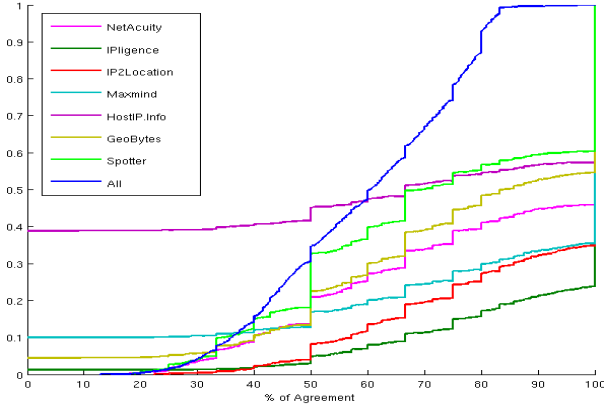}
\caption{CDF of Agreement Within Databases, 100km Radius}
\label{fig:cdf_agreement_100k}
\end{minipage}
\end{figure}

\begin{figure}
\begin{minipage}[b]{\linewidth}
\centering
\includegraphics[width=8cm, height=5.5cm]{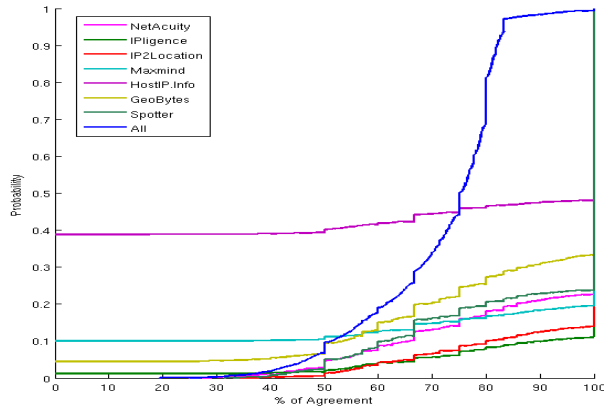}
\caption{CDF of Agreement Within Databases, 500km Radius}
\label{fig:cdf_agreement}
\end{minipage}
\end{figure}

Figures \ref{fig:cdf_agreement_100k} and \ref{fig:cdf_agreement}
present a CDF of the agreement within databases without singletons.
The X axis marks the percentage of IP addresses in PoPs that
represent the majority, and the Y axis presents the probability for
this majority vote. For Figure \ref{fig:cdf_agreement_100k} we set a
radius of 100km and in Figure \ref{fig:cdf_agreement} the used
radius is 500km, within which a majority is required. In some cases
no majority is found, i.e., less than 50\% of the IP addresses are
within any circle with the given radius. Remember that the algorithm
selects in such a case the location based on the largest group of
votes.

Note that for all databases there are PoPs that had no majority
vote, meaning the locations diverged by more than 100km or 500km.
IPLigence and IP2Location have the highest probability to reach an
agreement within a PoP, while HostIP.Info, and Geobytes grow at the
slowest pace. For a radius of $100km$, Spotter does not reach full
agreement for almost 60\% of the PoPs, probably due to measurement
accuracy limitations. Interestingly, for less than 4\% of the PoPs
there is $100\%$ agreement by all databases, which once again does
not correlate with single-database observations and points to a
mismatch between databases.

\subsection{Comparison Between Databases} \label{subsec:compare}
\subsubsection{Accuracy}
So far, we have discussed results that depend only on the database
itself. Next we compare the databases based on the data collected
from all databases. First, we asses the accuracy of a database by
comparing an IP location in every database to the location of its
PoP as voted by all databases.

\begin{figure}
\begin{minipage}[b]{\linewidth}
\centering
\includegraphics[width=8cm, height=5.5cm]{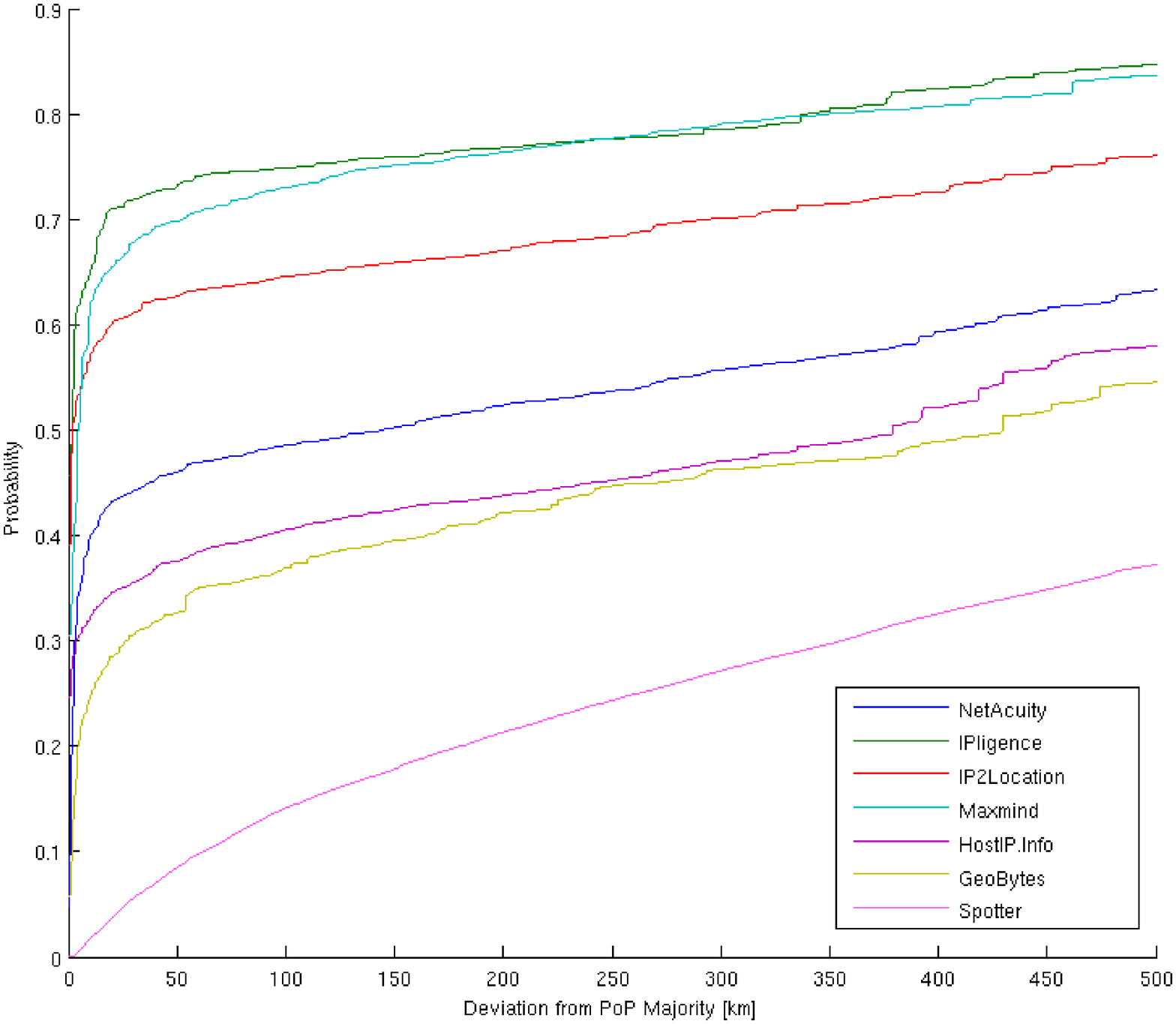}
\caption{CDF of database location deviation from PoP majority -
500km range} \label{fig:cdf_deviation_zoom}
\end{minipage}
\end{figure}

\begin{figure}
\begin{minipage}[b]{\linewidth}
\centering
\includegraphics[width=8cm, height=5.5cm]{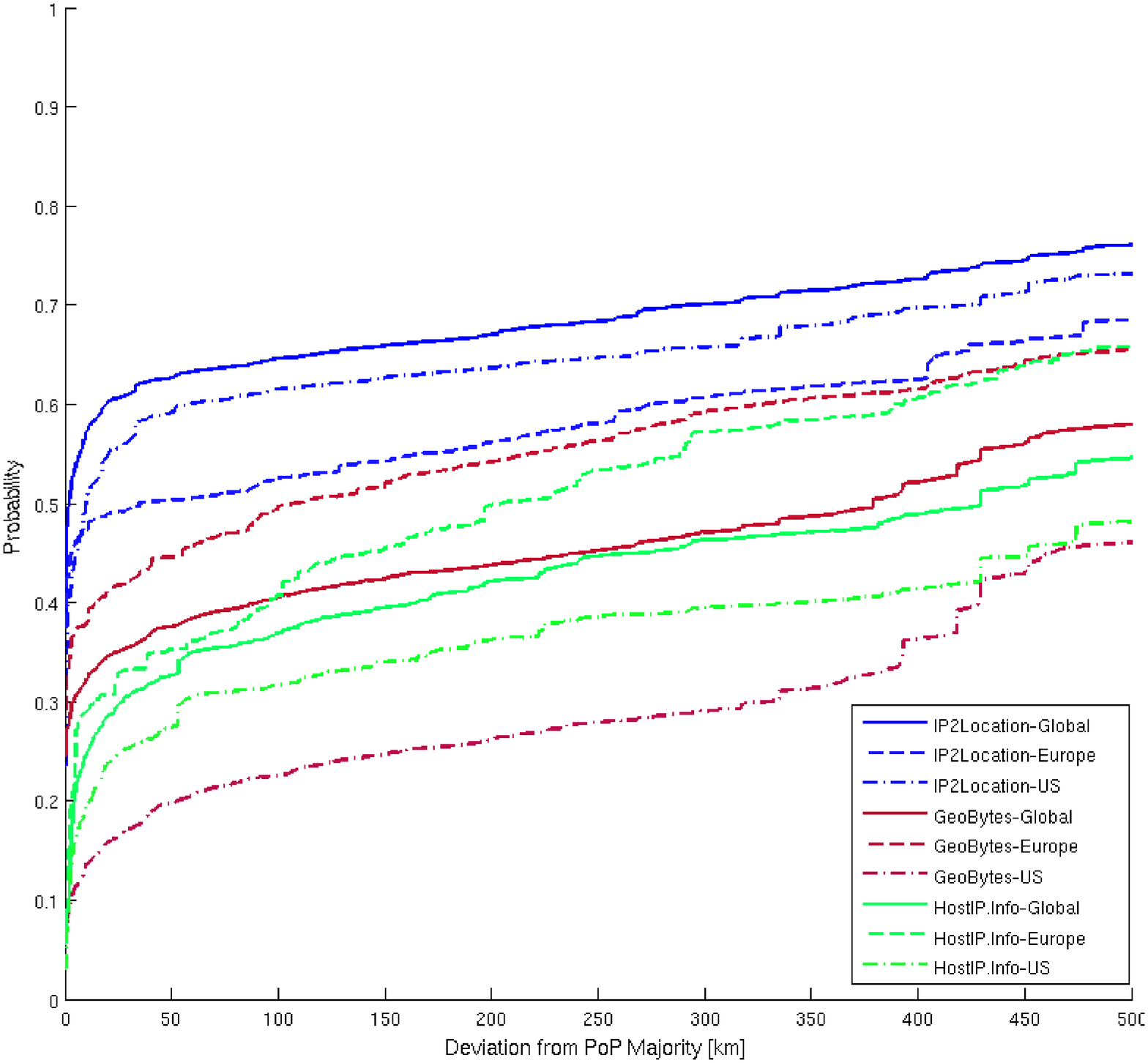}
\caption{Breakdown of deviation from PoP majority CDF By region -
500km Range} \label{fig:cdf_deviation_region}
\end{minipage}
\end{figure}

\begin{figure}
\begin{minipage}[b]{\linewidth}
\centering
\includegraphics[width=8cm, height=5.5cm]{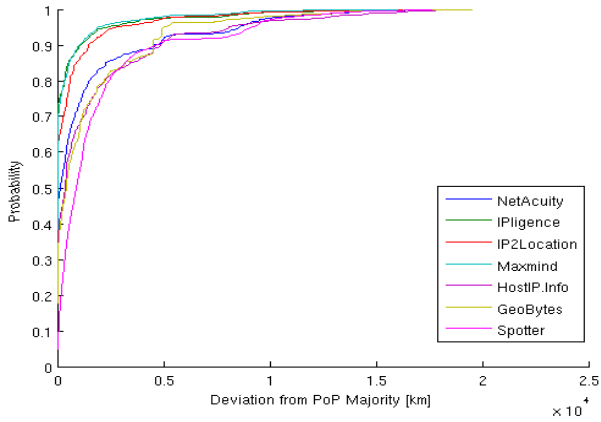}
\caption{CDF of Database location deviation from PoP majority}
\label{fig:cdf_deviation}
\end{minipage}
\end{figure}

Figure \ref{fig:cdf_deviation_zoom} depicts for each database the
CDF of the deviation of each IP from the PoP majority vote. The
interesting observations here consider $40km$ range, which is a city
range, and $500km$ range, which can be referred to as a region.
IPligence, MaxMind and IP2Location have a probability of 62\% to
73\% to place a IP within $40km$ from the PoP majority vote, with
IPligence and MaxMind placing over 80\% of the IP addresses within
$500km$ radius. Geobytes, HostIP.Info and Netacuity place 33\% to
47\% of the IP addresses within a city range, and 48\% to almost
60\% within $500km$ from the majority. Spotter places only 10\%
within $40km$ range and 30\% within the same region.

Some of the databases, like HostIP.Info, Netacuity, Geobytes and
Spotter, deviate less in Europe than in the USA and the rest of the
world, as depicted in Figure \ref{fig:cdf_deviation_region}. Other
databases, as IP2Location, have greater deviation in Europe than the
rest of the world. For clarity, only two of the databases are shown
in Figure \ref{fig:cdf_deviation_region}. A drawback of all
databases is that there is a long tail of IP addresses locations
which are placed $5000km$ or more from the majority of the vote.
Figure \ref{fig:cdf_deviation} shows that in some databases this
tail can hold 15\% of the IP addresses. Although the majority vote
may be incorrect, this points that at least one of the databases is
very far off from the real IP address location.

\begin{figure}
\begin{minipage}[b]{\linewidth}
\centering
\includegraphics[width=8cm,height=9cm]{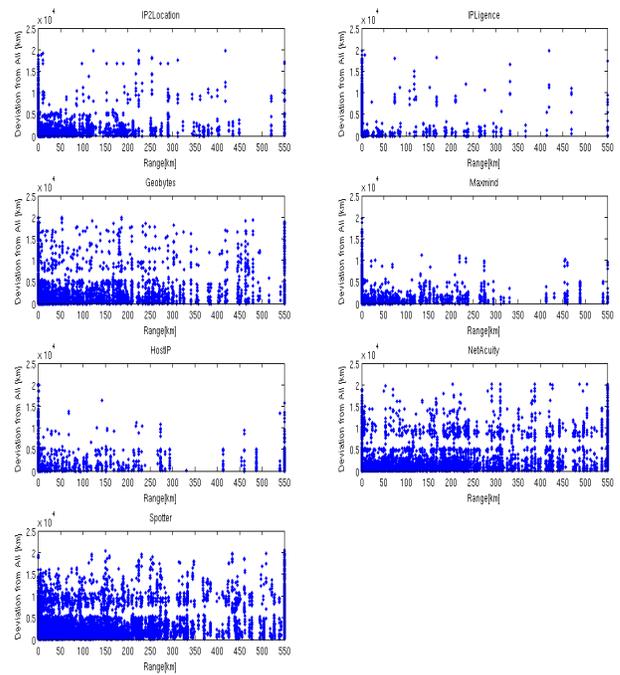}
\caption{Database location deviation from PoP majority vs. Range of
Convergance} \label{fig:range2deviation}
\end{minipage}
\end{figure}

Figure~\ref{fig:range2deviation} depicts for each database a scatter
plot of the range of convergence (X axis) versus the deviation of
the IP location from its PoP location based on all databases (Y
axis). The figure demonstrates that in many cases the range of
convergence is small , yet the deviation from the PoP majority vote
may be thousands of kilometers. Further more, a large range of
convergence does not imply that that the PoP center is necessarily
wrong, as again in all databases we see cases where the range is
large, yet the selected IP address location is the same as the
majority location from all databases. IPligence and IP2Location
demonstrate an interesting phenomenon: though their range of
convergence is very low, the variation from the PoP majority
location is very large. This can indicate, as is demonstrated next,
that large groups of IP addresses are assigned a single false
location.

For MaxMind and HostIP there are many PoPs at the far end of the
graph, with a large range of convergence. This is caused by lack of
information on specific IP addresses which does not allow them to
reach a majority vote. Netacuity and Spotter demonstrate a scattered
behavior, meaning the range of convergence and the deviation from
the PoPs majority both change. For Netacuity this means that IP
addresses are assigned distinct locations within the same area, as
with different users in the same city. Spotter suffers from large
range of convergence for some PoPs due to NULL replies, however
there is an obvious trend that places most PoPs IP addresses within
$300km$ range from each other, with a small number scattered at
larger range of convergance, as can be expected in a triangulation
based method.

\subsubsection{Correlation Between Databases}

\begin{figure}
\begin{minipage}[b]{\linewidth}
\centering
\includegraphics[width=8.5cm]{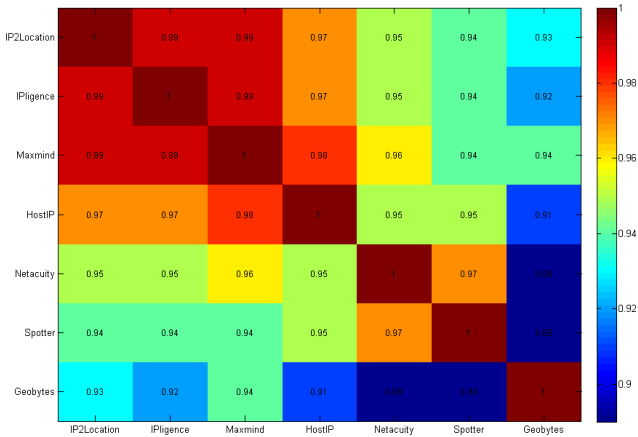}
\caption{Cross Correlation Between Databases - Heatmap}
\label{fig:heatmap}
\end{minipage}
\end{figure}

While some of the databases have proprietary means to gather
location information, a large portion of geolocation databases is
likely to come from the same source, such as getting country
information from ARIN. To examine this theory we calculate the cross
correlation between every pair of databases, on the entire IP
address location vector, and display it as a heatmap, shown in
Figure \ref{fig:heatmap}.

The strongest correlation is between IPligence and IP2Location. As
was shown in previous results, the trends of these two databases
look very similar. The correlation between these two databases is
over $0.99$. Maxmind and HostIP.Info also have very high correlation
with IP2Location and IPligence as well as between themselves. The
correlation figures above do not take into account NULL replies.
Considering those, IPligence's and IP2Location's correlation with
Maxmind drops to $0.8$ and with HostIP.Info below $0.6$. Comparing
all the databases, Netacuity and Geobytes correlate the least:
$0.89$. Spotter has over $0.94$ correlation with most databases,
expect Geobytes, with $0.97$ correlation to Netacuity. Considering
that the location given by Spotter is never a landmark, rather a
result of delay measurement, this is a high figure. The high
correlation between the databases indicates that in most cases the
location addresses returned by all the databases will be very much
alike. In cases where it is difficult to obtain the location
address, the answers may vary significantly between services.

\subsection{Database Anomalies}
Though the results above may indicate that some databases have
superb location information, this is not the case. In many cases the
returned data is deceiving, and actually may represent lack of
information in the database. For example, we identified 266 IP
addresses in the PoPs that belong to Qwest Communications. Out of
those, 253 IP addresses are located by IPligence in Denver,
Colorado. Looking at the raw IPligence database, there are 20291
entries that belong to Qwest communications. Out of those, $20252$
are located in Denver, which is the location of Qwest's
headquarters. The phenomenon was first detected by our algorithm
last year, in July/2009: 70 Qwest PoPs where detected. Maxmind
assigned them to 55 different locations, HostIP.info to 46
locations, IP2locations to 35 locations and IPligence located them
all in Denver. In response to a query back then, IPligence have
replied that "In some occasions you could find records belonging to
RIPE or any other registrar, these are most likely not used IP
addresses but registered under their name, anything else should be
empty or null".

Quite a similar case exists with IP2Location. For Cogent, 2365 out
of 2879 IP addresses were located in Washington DC, which is
Cogent's headquarters location. Out of 57 PoPs belonging to Cogent,
only one was not placed by IP2Location in these exact same
coordinates. For IPligence, all the PoPs were located in the same
place, too. However, Maxmind placed the PoPs in 13 locations,
Geobytes in 23 locations and Netacuity in 31 locations (only a
handful in Washington's area).
In the Akamai audit by Gomez~\cite{akamai_audit} a similar case
is described: A node in Vancouver, Canada was reported to be in
Tornto, and a node in Bangalore, India was reported to be in Mumbai.
In both cases those were ISP headquarters known locations.

\begin{figure}
\begin{minipage}[b]{\linewidth}
\centering
\includegraphics[width=8cm,height=5cm]{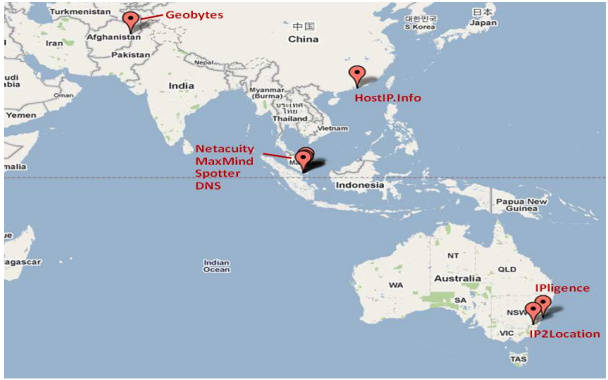}
\caption{Mismatch Between Databases - UUNET }
\label{fig:uunet_mismatch}
\end{minipage}
\end{figure}

Sometimes differences between databases may be very acute, with a
reported node location being far off by thousands of kilometers and
even countries far apart. In Figure \ref{fig:uunet_mismatch} one
such example is shown. We take a 4-nodes PoP in ASN 703 (Verizon/
UUNET / MCI Communications) and display on a map the location of the
PoP based on each of the geolocation database. IPligence,
IP2Location, Geobytes, Netacuity and DNS all internally have the PoP
four IP addresses at the same location, however each of the
databases locate it differently: IPligence and IP2Location in
Australia, Netacuity and DNS in Singapore and Geobytes in
Afghanistan. MaxMind and Spotter lack information on these nodes and
HostIP.Info places the PoP with 66\% certainty in China. Extending
our PoP view to include singletons, thus including 10 nodes, the
picture does not change.  MaxMind and Spotter have location on one
of the IPs and they place it in Singapore. IPligence and IP2location
place 9 out of 10 IPs in Australia, and one in Singapore. Geobytes
places this last IP address in Singapore too, yet 6 out of 10 IP
locations still point to Kabul. The rest three nodes are located in
Australia. Geobytes does give low certainty rate to the location,
being 50 or less to both country and region. Netacuity places 8 out
of 10 IPs in Singpore and 2 in Australia. HostIP.Info has location
information on 6 IPs, 3 of them are placed in China and 3 in
Australia, but in Melbourne, far from IPligence and IP2location
designated location. Notably, all the edges in this PoP have less
than 3.5mS delay and are measured five to 173 times each.

\begin{figure}
\begin{minipage}[b]{\linewidth}
\centering
\includegraphics[width=8cm]{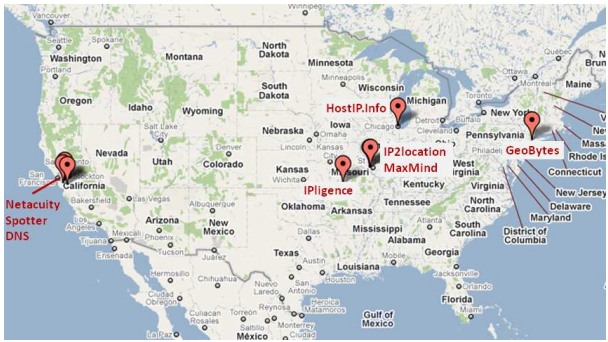}
\caption{Mismatch Between Databases - Global Crossing}
\label{fig:gcrossing_mismatch}
\end{minipage}
\end{figure}

The mismatch between databases is not uncommon. Some examples exist
inside the United States, too: in Figure
\ref{fig:gcrossing_mismatch} we show one PoP in ASN 3549, Global
Crossing, as it is placed by the different geolocation databases all
across the country. This PoP has over 160 IP addresses, counting
singletons, and as such a majority in each database has more
substance. IPligence places the PoP with more than 90\% majority in
Springfield, Missouri. MaxMind and IP2Location point to Saint Louis,
Missouri with 92\% and 82\% accordingly. NetAcuity indicates that
the PoP is in  San-Jose, California with 100\% certainty, while DNS
and Spotter place the PoP in this vicinity, in a radius of a few
tens of kilometers. GeoBytes has somewhat above 59\% of the
locations pointing to New York, with other common answers being
spread across California (25\%). Geobytes country certainty here was
100\% with 42\% region certainty for the IP addresses it located in
New York. HostIP.Info placed the PoP in Chicago with 65\% majority
(28\% of the locations had pointed to Santa Clara, California).

The above are not single incidents. Similar cases have been found in
other AS as well, such as REACH (AS 4637), where IPligence,
IP2location and Maxmind located a PoP in China, Geobytes located it
in Australia, while Netacuity and Spotter put it in the silicon
valley, USA. Other cases range from AS16735 (CTBC/Algar Telecom)
where PoP locations in Brazil were set thousands of kilometers
apart, to Savvis (AS3561) which is another case of locations spread
across the USA.

\subsection{Database Changes}
One of the motivations to update geolocation databases is the claim
that they change significantly over time. Maxmind\cite{maxmind}
claim that it looses accuracy at a rate of approximately 1.5\% per
month. IP2Location~\cite{ip2location} state that on average, there
are 5\%-10\% of the records being updated in the databases every
month due to IP address range relocation and new range available.
Based on the PoPs dataset, we compare this information versus the
databases at our disposal. For IPligence, an average of
approximately one percent of the addresses changes every month, with
some minimal changes in some consecutive months, such as 0.6\%
between November and December 2009. In HostIP.Info, 18\% of the IP
addresses changed their location within nine months, meaning an
average of 2\% a month. IP2Location changed only 1\% of the
locations over 4 months, meaning 0.25\% per month, however the
reference set here included only 10K IP addresses. For Netacuity,
running only on our dataset of 104K IP addresses, we observe that
2.4\% of the IP addresses have changed in less than a month.

\section{Discussion}\label{discussion.sec}

Interpretation of the results in Section \ref{subsec:compare} should
be done with care.  Placing a PoP at the majority center of gravity
may not always yield the true results, e.g., in cases where a single
wrong information source is used by multiple databases.

IPligence and IP2Location share several similar characteristics as
well as strong cross correlation. This is exhibited by the high
probability for a small range of convergence and by the fast rate
their probability grows to reach a high level of agreement. However,
the various anomalies found in their databases shed a different
light on these results. For example, if for a certain ISP all the IP
addresses are assigned to a single location, then the immediate
effect will be a small range of convergence and high level of
agreement. Further more, lack of NULL replies in this case may be
misleading as the returned reply may be false. As in the cases of
MCI/UUNET and Global Crossing, as well as other investigated cases,
IP2Location and IPligence located IP addresses far from Spotter's
estimated location, it is likely that their geolocation information,
in these cases, was wrong.

Judging MaxMind performance has to be done carefully, as they do not
claim to have high accuracy for router interfaces. This is
manifested in the high number of returned NULL replies. MaxMind
seems to have a lot in common with IPligence and IP2Location, as the
cross correlation shows, however unlike these databases MaxMind
prefers to return NULL or country center reply when a location is
unknown and thus returns less false locations.

We find it hard to analyze Geobytes performance. The database
returns relatively a lot of NULL replies, which are significantly
reduced by PoP level aggregation. Further more, the granularity of
the database is /24 CIDR, thus grouping every 256 address block to a
single location. Though this is a common practice, it has some
degradation effect. The oddity here is that there are many cases
where the range of convergence is about 250km, which means the
database located the IP addresses within each others' area, but did
not consider them to be at the exact same city. However, as Geobytes
indicate that they focus on the service area of a PoP rather then
the location of that PoP, this can be expected. An advantage of
Geobytes is that no assignment to a single location or ISP
headquarters were detected so far.

The main drawback of HostIP.Info is the lack of information. As most
of the IP addresses and over a third of the PoPs did not have any
location information, the knowledge gained on this database is
limited. The limited location information led HostIP.info to perform
worst on almost all test cases. In addition, in more than a single
case the location information indicated by the database was far off
from any other database or measurement based location.

Netacuity is probably the most expensive and highly claimed database
used in this research. The results of the tests however may not
stand up to expectations. Though one may assume that majority
location is affected by errors in other databases, it can be
expected that when compared to itself the performance will be high.
The results show that for over 40\% of the cases, same PoP IP
addresses are not all located within 100km radius, which is in fact
200km diameter, and close to 20\% are not located within 500km
radius either. The strength of Netacuity is that ISP IP addresses
are rarely assigned to a single location, unless this is indeed a
true single place. In addition, in the several anomalous cases that
were investigated, Netacuity majority pointed to the most probable
correct location. Note that a minority of IP location votes still
pointed to different locations, even in different countries.

\subsection{Active Measurement Accuracy}

Active measurements are used by many geolocation services
\cite{Katz06,WongSS07,spotter} and by other projects for different
localization tasks, most notably for assigning IP addresses to PoPs
\cite{Spring02}. Spotter geolocation is based solely on active
measurements, thus we selected to study its performance to a greater
depth due to the importance of understanding the limitations of this
approach.

\begin{figure}
\begin{minipage}[b]{\linewidth}
\centering
\includegraphics[width=8cm,height=4cm]{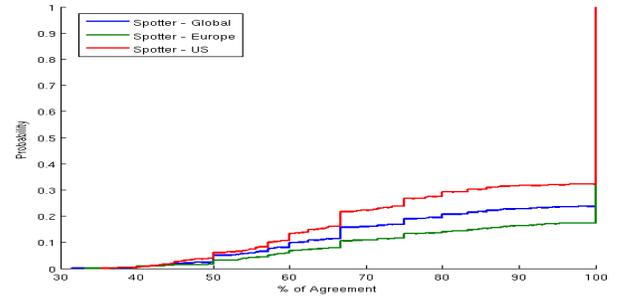}
\caption{Breakdown of the agreement CDF for Spotter by region.} \label{fig:spotter-agree}
\end{minipage}
\end{figure}

\begin{figure}
\begin{minipage}[b]{\linewidth}
\centering
\includegraphics[width=8cm,height=4cm]{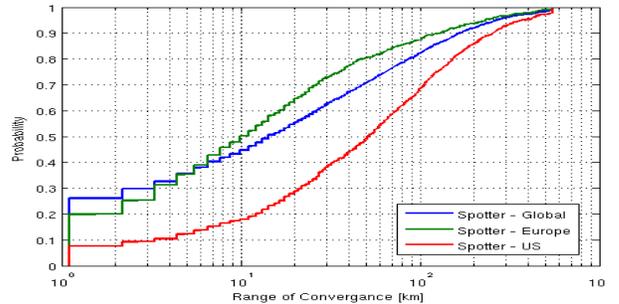}
\caption{Breakdown of the convergence range CDF for Spotter by
region.} \label{fig:spotter-range}
\end{minipage}
\end{figure}

\begin{figure}
\begin{minipage}[b]{\linewidth}
\centering
\includegraphics[width=8cm,height=5cm]{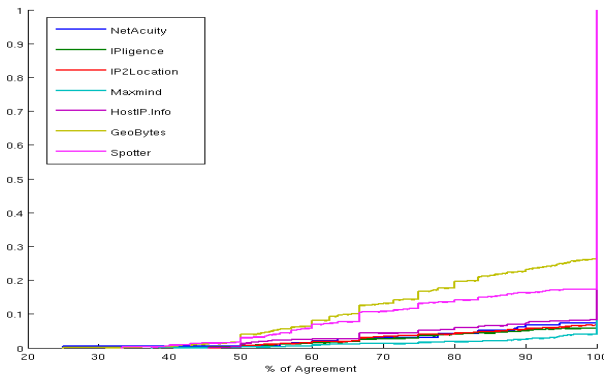}
\caption{CDF of Agreement Within Databases In Europe, 500km Radius}
\label{fig:agree-EU}
\end{minipage}
\end{figure}

Figures \ref{fig:spotter-agree} and \ref{fig:spotter-range} show
Spotter's overall performance compared with its performance for PoPs
located only in Europe or in the USA. It is clear from both figures
that in Europe Spotter perform much better than in the USA and
slightly better than the world average.  For example, for 40km
radius (which is frequently used as a city diameter) Spotter reach
about 78\% convergence in Europe compared to 67\% convergence
worldwide, and only 44\% for the USA. The difference can be
explained\footnote{We consulted Peter Haga and Peter Matray from the
Spotter project on this aspect.} by the spread of vantage points
used by Spotter, which are almost entirely based on PlanetLab nodes.
While in Europe PlanetLab nodes are well spread geographically, in
the USA, most PlanetLab nodes are located along the coasts making
localization of IP addresses in the middle of the USA less accurate.
Interestingly, other databases which are based on other means also
achieve better results for European addresses than for USA addresses
(see Fig.~\ref{fig:agree-EU}).

Spotter convergence (Fig.~\ref{fig:cdf_range}) starts as the lowest
which is an outcome of the measurement error that tend to spread the
results for different IPs around the `true' location.  However, at a
radius of 100km it closes the gap with most databases and reaches
over 80\% convergence (and close to 90\% for Europe). However, 20\%
`error' may make distance measurements unfit as the sole method for
assigning IP address to PoPs.

\section{Conclusion}\label{conclusion.sec}

This paper presented a comprehensive study of geolocation databases,
comparing a large number of databases of different types. The
results show that while some of the databases provide results that
are well aggregated and have a small number of NULL replies, the
accuracy of the returned location can not be trusted. There is a
strong correlation between all databases, which indicated that the
vast majority of location information replies are correct. However,
in some cases there are errors in the databases in the range of
thousands of kilometers and countries apart. The use of geolocation
database should therefore be careful and its information can not be
considered as ground truth.

Our results also show that measurement based geolocation can achieve
good results that compete with geolocation information gathered by
other means and that the achieved accuracy of geolocation using such
tools can be fairly high. However, this accuracy may not be high
enough to be used as a the sole tool to map IP addresses to PoPs.
Future research in this field should focus on means to decide on
ground truth when there is a disagreement between the databases.

\section{Acknowledgements}
We would like to thank Peter M\'{a}tray and Peter H\'{a}ga, as well
as other members of the Spotter project\cite{spotter}\, for
providing us with their dataset and helping this work. We would also
like to thank Frank Bobo from Digital Envoy, Adrian McElligott from
Geobytes and Edward Lin from Maxmind for helping us in obtaining
their databases and answering our questions.

\bibliographystyle{abbrv}
\bibliography{cites}

\begin{thebibliography}{10}

\bibitem{ipinfo}
{IPInfoDB}.
\newblock http://ipinfodb.com, 2010.

\bibitem{quova}
{Quova}.
\newblock http://www.quova.com, 2010.

\bibitem{akamai}
{Akamai}.
\newblock {EdgePlatform}.
\newblock $http://www.akamai.com/html/technology/$\\$edgeplatform.html$, 2010.

\bibitem{att_map}
ATT-Global-Services.
\newblock Att global services global network map.
\newblock http://www.corp.att.com/globalnetworking/\\media/ network\_map.swf.

\bibitem{british_map}
BT-Global-Services.
\newblock Network maps.
\newblock http://www.bt.net/info/europe.shtml.

\bibitem{netacuity}
{Digital Envoy}.
\newblock {NetAcuity Edge}.
\newblock $http://www.digital-element.com/our\_technology/edge.html$, 2010.

\bibitem{EBSN10}
B.~Eriksson, P.~Barford, J.~Sommers, and R.~Nowak.
\newblock A learning-based approach for ip geolocation.
\newblock In {\em Passive and Active Measurement}, pages 171--180, 2010.

\bibitem{FeldmanS08}
D.~Feldman and Y.~Shavitt.
\newblock Automatic large scale generation of internet {PoP} level maps.
\newblock In {\em GLOBECOM}, pages 2426--2431, 2008.

\bibitem{geobytes}
{Geobytes}.
\newblock {GeoNetMap}.
\newblock $https://secure.geobytes.com/GeoNetMap.htm$, 2010.

\bibitem{gblx_map}
Global-Crossing.
\newblock {Global Crossing Network}.
\newblock http://www.globalcrossing.com/html/ map062408.html.

\bibitem{GS07}
G.~Goodell and P.~Syverson.
\newblock The right place at the right time.
\newblock {\em Commun. ACM}, 50(5):113--117, 2007.

\bibitem{google}
{Google}.
\newblock {Geolocation API}.
\newblock $http://code.google.com/apis/gears/$\\$api\_geolocation.html$, 2010.

\bibitem{google2}
{Google Gears Wiki}.
\newblock {Geolocation API}.
\newblock $http://code.google.com/p/gears/wiki/GeolocationAPI$, 2010.

\bibitem{CiscoPoP}
B.~R. Greene and P.~Smith.
\newblock {\em Cisco ISP Essentials}.
\newblock Cisco Press, 2002.

\bibitem{GUF07}
B.~Gueye, S.~Uhlig, and S.~Fdida.
\newblock Investigating the imprecision of ip block-based geolocation.
\newblock In {\em PAM'07: Proceedings of the 8th international conference on
  Passive and active network measurement}, pages 237--240, 2007.

\bibitem{GZCF06}
B.~Gueye, A.~Ziviani, M.~Crovella, and S.~Fdida.
\newblock Constraint-based geolocation of internet hosts.
\newblock {\em IEEE/ACM Trans. Netw.}, 14(6):1219--1232, 2006.

\bibitem{ip2location}
{Hexsoft Development}.
\newblock {IP2Location}.
\newblock http://www.ip2location.com, 2010.

\bibitem{hostip}
{hostip.Info}.
\newblock {hostip.info}.
\newblock http://www.hostip.info, 2010.

\bibitem{IETF-Geo}
IETF.
\newblock Geopriv workgroup.
\newblock {http://tools.ietf.org/wg/geopriv/}.

\bibitem{akamai_audit}
G.~Inc.
\newblock Akamai ip location service performance assessment, akam01, September
  2009.

\bibitem{ipligence}
{IPligence}.
\newblock {IPligence Max}.
\newblock http://www.ipligence.com, 2010.

\bibitem{JE06}
J.~M. Jensen.
\newblock Personal jurisdiction in federal courts over international e-commerce
  cases.
\newblock {\em Loyola of Los Angeles Law Review}, 1507, 2006.

\bibitem{KF10}
E.~Kaiser and W.-c. Feng.
\newblock Helping ticketmaster: Changing the economics of ticket robots with
  geographic proof-of-work.
\newblock In {\em Proceedings of Global Internet 2010}, March 2010.

\bibitem{Katz06}
E.~Katz-Bassett, J.~P. John, A.~Krishnamurthy, D.~Wetherall, T.~Anderson, and
  Y.~Chawathe.
\newblock Towards {IP} geolocation using delay and topology measurements.
\newblock In {\em The 6th ACM SIGCOMM conference on Internet measurement
  (IMC'06)}, pages 71--84, 2006.

\bibitem{King09}
K.~F. King.
\newblock Geolocation and federalism on the internet: Cutting internet
  gambling's gordian knot.
\newblock {\em Columbia Science and Technology Law Review}, Forthcoming, 2009.

\bibitem{quova_audit}
P.~C. LLP.
\newblock Quova - report of independant accountants.
\newblock
  $http://www.quova.com/documents/$\\$PricewaterhouseCoopers\_Audit.pdf$,
  October 2008.

\bibitem{Ma09}
S.~Malphrus.
\newblock Perspectives on retail payments fraud.
\newblock {\em Economic Perspectives}, XXXIII(1).

\bibitem{maxmind_city}
{MaxMind LLC}.
\newblock {GeoIP City Accuracy for Selected Countries}.
\newblock $http://www.maxmind.com/$\\$app/city\_accuracy$, 2008.

\bibitem{maxmind}
{MaxMind LLC}.
\newblock {GeoIP}.
\newblock http://www.maxmind.com, 2010.

\bibitem{geobytes_patent}
A.~E. McElligott.
\newblock Method and software product for identifying network devices having a
  common geographical locale, May 2007.

\bibitem{MO09}
J.~A. Muir and P.~C.~V. Oorschot.
\newblock Internet geolocation: Evasion and counterevasion.
\newblock {\em ACM Comput. Surv.}, 42(1):1--23, 2009.

\bibitem{PS01}
V.~N. Padmanabhan and L.~Subramanian.
\newblock An investigation of geographic mapping techniques for internet hosts.
\newblock In {\em SIGCOMM '01: Proceedings of the 2001 conference on
  Applications, technologies, architectures, and protocols for computer
  communications}, pages 173--185, 2001.

\bibitem{qwest_map}
Qwest.
\newblock {IP Network Statistics}.
\newblock http://66.77.32.148/index\_flash.html.

\bibitem{LMHCV09}
P.~H. I. C. G.~V. S.~Laki, P.~M\'{a}tray.
\newblock A model based approach for improving router geolocation.
\newblock {\em Computer Networks}, 54(9):1490--1501, 2010.

\bibitem{spotter}
{S. Laki, P. M\'{a}tray, P. H\'{a}ga, T. Seb\"{o}k, I. Csabai, G. Vattay}.
\newblock Spotter: A model based active geolocation tool.
\newblock (to be published), 2010.

\bibitem{JUNIPOP}
A.~Sardella.
\newblock Building next-gen points of presence, cost-effective pop
  consolidation with juniper routers.
\newblock White paper, Juniper Networks, June 2006.

\bibitem{DHSCS09}
H.~Security.
\newblock A roadmap for cyber security research.
\newblock http://www.cyber.st.dhs.gov/docs/DHS-Cybersecurity-Roadmap.pdf,
  November 2009.

\bibitem{DimesSigcomm05}
Y.~Shavitt and E.~Shir.
\newblock {DIMES}: Let the internet measure itself.
\newblock In {\em ACM SIGCOMM Computer Communication Review}, volume~35, Oct.
  2005.

\bibitem{SZ-Netsci10}
Y.~Shavitt and N.~Zilberman.
\newblock A structural approach for pop geo-location.
\newblock In {\em Infocom Workshop on Network Science for Communications
  (NetSciCom)}, March 2010.

\bibitem{SGU08}
S.~S. Siwpersad, B.~Gueye, and S.~Uhlig.
\newblock Assessing the geographic resolution of exhaustive tabulation for
  geolocating internet hosts.
\newblock In {\em Passive and Active Measurement}, volume 4979, pages 11--20,
  2008.

\bibitem{Spring02}
N.~Spring, R.~Mahajan, and D.~Wetherall.
\newblock Measuring isp topologies with rocketfuel.
\newblock In {\em In Proc. ACM SIGCOMM}, pages 133--145, 2002.

\bibitem{sprint_map}
Sprint.
\newblock {Global IP Network}.
\newblock https://www.sprint.net/network\_maps.php.

\bibitem{SV07}
D.~Svantesson.
\newblock E-commerce tax: How the taxman brought geography to the 'borderless'
  internet.
\newblock {\em Revenue Law Journal}, 17.1, 2007.

\bibitem{SV08}
D.~Svantesson.
\newblock How does the accuracy of geo-location technologies affect the law?
\newblock {\em Law papers}, 2008.

\bibitem{IETF-LIS}
M.~Thomson and J.~Winterbottom.
\newblock Discovering the local location information server (lis).
\newblock http://tools.ietf.org/html/draft-ietf-geopriv-lis-discovery-15, March
  2010.

\bibitem{WongSS07}
B.~Wong, I.~Stoyanov, and E.~G. Sirer.
\newblock Octant: A comprehensive framework for the geolocalization of internet
  hosts.
\newblock In {\em NSDI}, 2007.

\bibitem{Yoshida09}
K.~Yoshida, Y.~Kikuchi, M.~Yamamoto, Y.~Fujii, K.~Nagami, I.~Nakagawa, and
  H.~Esaki.
\newblock Inferring pop-level isp topology through end-to-end delay
  measurement.
\newblock In {\em PAM}, volume 5448, pages 35--44, 2009.

\bibitem{ZRPR06}
M.~Zhang, Y.~Ruan, V.~Pai, and J.~Rexford.
\newblock How dns misnaming distorts internet topology mapping.
\newblock In {\em ATEC '06: Proceedings of the annual conference on USENIX '06
  Annual Technical Conference}, pages 34--34, 2006.

\end{thebibliography}

\vfill

\end{document}